\documentclass[letterpaper,twocolumn,aps]{revtex4}
\usepackage[latin1]{inputenc}
\usepackage{latexsym,amstext,amssymb,amsmath}
\usepackage[dvips]{graphicx}
\usepackage{color}
\usepackage{xspace}

\newcommand{\SRO}{SrRuO$_3$\xspace}
\newcommand{\GFO}{GdFeO$_3$ }
\newcommand{\CTO}{CaTiO$_3$ }

\begin{document}

\title{
Structural, electronic and magnetic properties of \SRO under epitaxial strain
}

\author{A. T. Zayak$^{1}$}
\email[Electronic address:]{zayak@physics.rutgers.edu}

\author{X. Huang$^{2}$}

\author{J. B. Neaton$^{3}$}

\author{Karin M. Rabe$^{1}$}

\affiliation{$^{1}$Department of Physics and Astronomy, Rutgers University\\
Piscataway, New Jersey 08854-8019, USA}

\affiliation{$^{2}$
RJ Mears, LLC,
1100 Winter Street, Suite 4700,
Waltham, MA 02451 }

\affiliation{$^{3}$The Molecular Foundry, Material Sciences Devision,
  Lawrence Berkeley National Laboratory, Berkeley CA 94720, USA}

\date{\today}

\begin{abstract}
Using density functional theory within the local spin density
approximation, structural, electronic and magnetic properties of SRO
are investigated. 
We examine the magnitude of the orthorhombic distortion in the ground
state and also the effects of applying epitaxial
constraints, whereby the influence of large (in the range of $\pm
4\%$) in-plane strain resulting from coherent epitaxy, for both [001]
and [110] oriented films, have been
isolated and investigated.  The overall pattern of the structural
relaxations reveal coherent distortions of the oxygen octahedra
network, which determine stability of the magnetic
moment on the Ru ion. The structural and magnetic parameters exhibit
substantial changes allowing us to discuss the role of symmetry and
possibilities of magneto-structural tuning of \SRO-based thin film
structures.  
\end{abstract}


\maketitle

\section{introduction}

Epitaxial thin film technology has the potential to change the
functionality of conventional crystals in a way to achieve better
performance and novel properties. Grown layer by layer in thin film
geometries, the crystals still retain their bulk-like properties in
two in-plane lateral dimensions, while being strongly modified
along the direction of growth by finite-size effects, presence of
interfaces and surfaces, and the misfit strain from epitaxial
lattice-matching. Additional interactions which emerge between
consequent layers allow control, tuning and enhancing of their
functional properties.     
Complex artificially-designed heterostructures with specific
characteristics can be synthesized with atomic-level precision
\cite{ramesh:2002}.  

In many perovskite-oxide based heterostructures and superlattices, the
metallic ferromagnetic perovskite SrRuO$_3$ plays a particularly key
role, often serving as an electrode material that allows for better
integration and facilitates electrical measurements
\cite{maki:2003,ban:2004}. However, recent studies of epitaxial
thin films \cite{gan:1998} suggest that \SRO may have novel
magnetostructural properties in ultrathin film form. By tuning film
thickness and in-plane strain (through choice of substrate) very
different properties may emerge as compared with the bulk.
Similar approaches have been exploited in recent experimental and
theoretical investigations showing modifications of the ferroelectric
epitaxial thin films with strongly enhanced performance  \cite{
nagarajan:1999, Fuchs,Lee_nature,LG,armin:2004}. However, to our
knowledge, the impact of strain on ferromagnetic perovskite
films has not yet been studied.

\begin{figure}
\includegraphics[width=7.0cm]{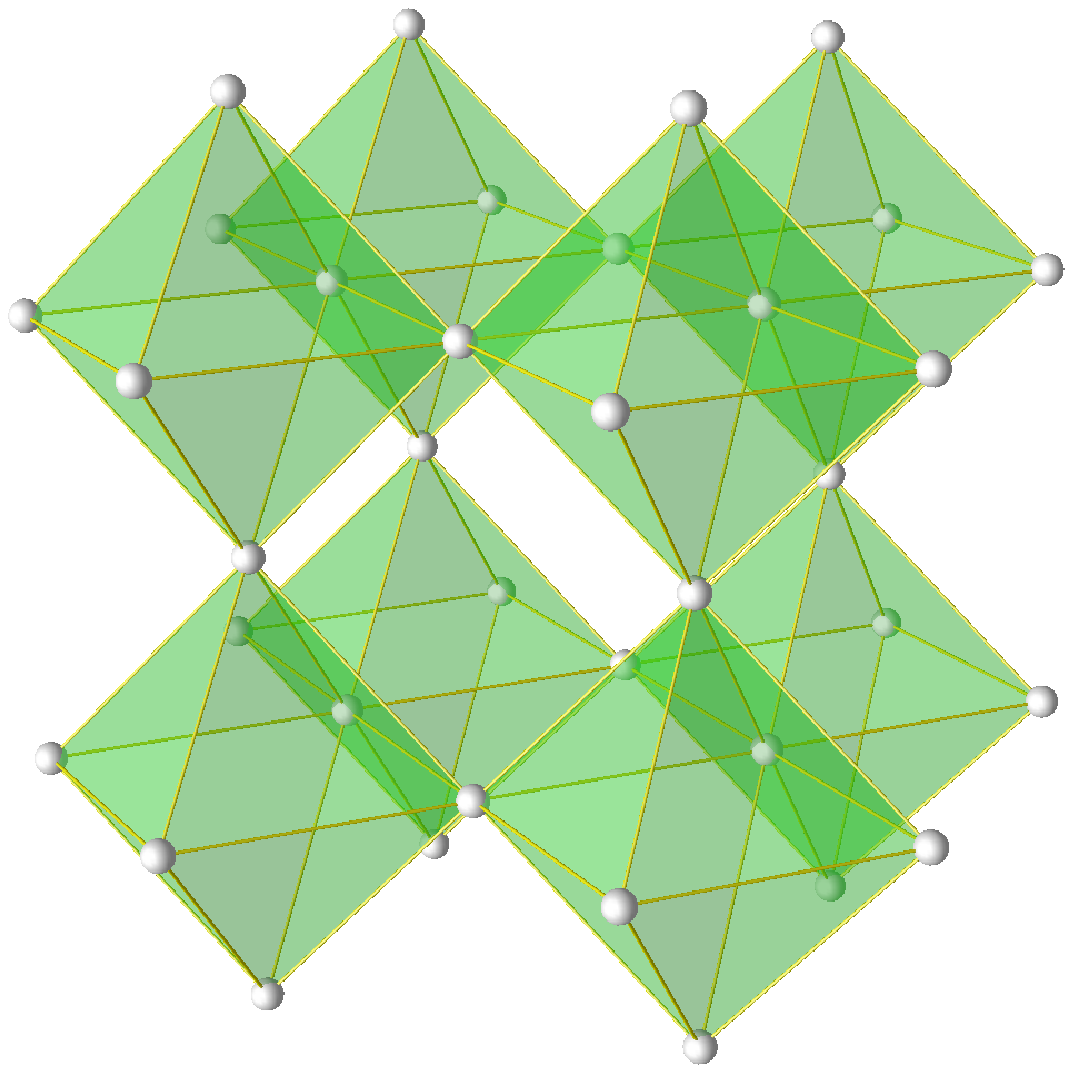}\\
\vspace{5mm}
\includegraphics[width=7.0cm]{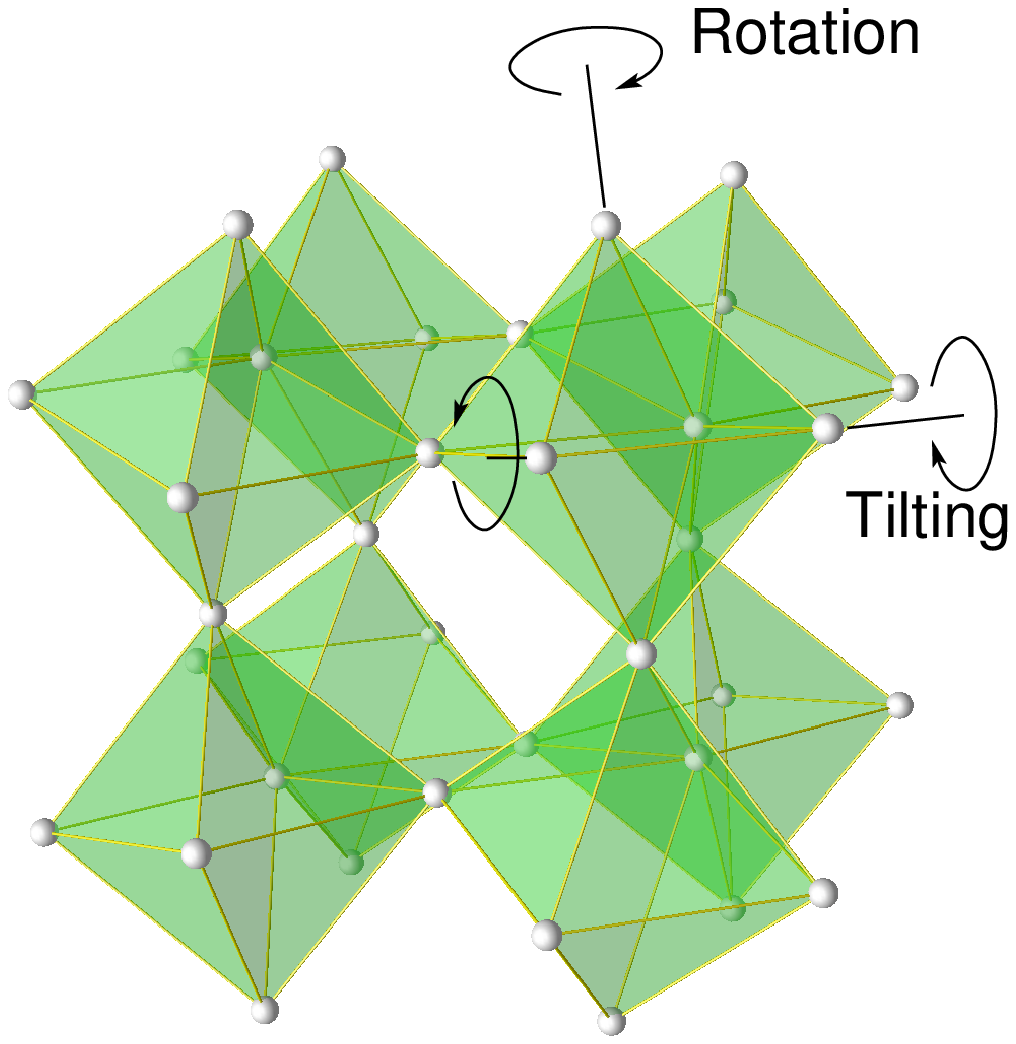}
  \caption{\label{octahedrons}  
(Top picture) 
The structure of the oxygen network in a $Pm\overline{3}m$
perovskite structure. The oxygen octahedra are perfectly regular.
 (Bottom picture)
The structure of orthorhombic $Pbnm$ perovskite structure with tilting
(rotation of the octahedral cages). The oxygen octahedra in this case
are not necessary regular. We consider this aspect in section III of
this paper. }
\end{figure}

Bulk \SRO undergoes a series of phase transformations with decreasing
temperature (for more details see Ref. \cite{chako:1998,Feizhou}).
It is a metallic ferromagnet with a Curie temperature about 160~K
and a magnetic moment of 1.1 $\mu_{B}$ per formula unit \cite{kanba:1976}.
The ideal cubic perovskite structure with cubic symmetry
$Pm\overline{3}m$ is stable above 950 K. From this temperature and
down to 820 K it appears in the tetragonal $I4/mcm$ structure. Below 850 K
and at standard pressures, \SRO has an orthorhombically distorted
perovskite structure (space group $Pbnm$ 
\footnote{According to crystallographic rules the notation of this
  space group is $Pnma$ which has the doubled $b$ axis with respect to the
  simple perovskite structure. However, the physics community
  traditionally prefers the $c$ axis to be the reference direction,
  therefore the $Pbnm$ is used in this work. These two notations are
  related via a simple transformation \cite{vanaken}. For the space
  group $Pnma$ we have lattice parameters: $a_1 \sim \sqrt{2}a_p$,
  $b_1 \sim 2a_p$, $c_1 \sim \sqrt{2}a_p$, where $a_p$ is the lattice
  parameter of the cubic perovskite structure. Space group $Pbnm$ has
  lattice parameters: $a_2 \sim \sqrt{2}a_p$,
  $b_2 \sim \sqrt{2}a_p$,  $c_2 \sim 2a_p$. The transformation from
  $Pnma$ to $Pbnm$ is: $a_1 \longrightarrow b_2$, $b_1 \longrightarrow
  c_2$, $c_1 \longrightarrow a_2$.} 
with 20 atoms per cell \cite{jones:1989}).  The structure is
the \GFO/\CTO-type, which can be understood by first doubling the 5-atom cell
along [110] with both oxygen octahedra being rotated in the same way
around [001] ($z$-axis), and then doubling this 10-atom
system along [001] with two oxygen octahedra connected along the
$z$-axis, being further tilted in opposite directions.  
This structure is shown in Fig~\ref{octahedrons}.
In the Glazer notations, the tilting of \SRO is described by
($a^-a^-c^+$) \cite{Woodward,Lufaso}. One of our main points below
will be that in \SRO, the rotation and tilting is accompanied by a
substantial deformation of the oxygen octahedra.

\begin{figure}[b]
\includegraphics*[angle=0,width=7.0cm]{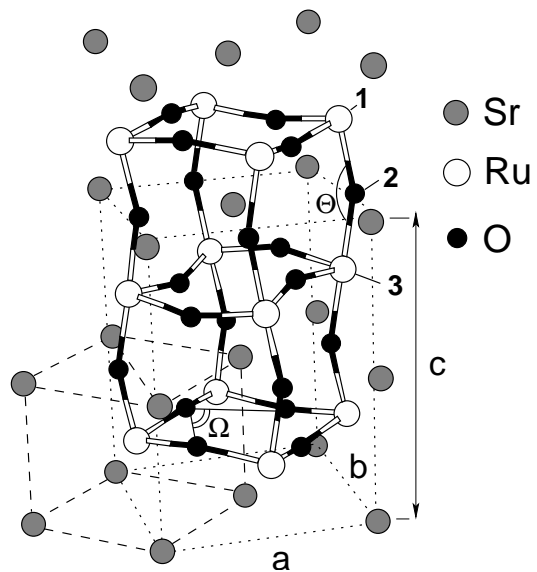}
  \caption{\label{structure}  
 The $Pbnm$ crystal structure of SrRuO$_3$
  (\GFO/\CTO-type). The unit cell consists of four formula units of the ideal
  cubic perovskite structure. The Ru atoms possess relatively large
  magnetic moments 
  and occupy high-symmetry positions in the orthorhombic unit
  cell. The atoms of  oxygen and Sr are displaced from their
  high-symmetry 
  positions due to the tilting. The numbers attached to certain atoms
  help us to define the angle of tilting as discussed in the
  section  ``Methods''.}    
\end{figure}

Previous first-principles studies of \SRO examined 
the electronic and magnetic properties of the bulk
low-temperature $Pbnm$ structure
\cite{singh:1996,allen:1996,MazinSRO}. Their results 
largely agree with experimental observations and explain
basic features of the ground state in the absence of strain.
In this paper, we use first-principles calculations to isolate and
investigate the effect of epitaxial strain on the structure of \SRO.  
We explore \SRO with this constraint through examination of the
orthorhombic distortion, magnetic and electronic properties of the
$Pbnm$ structure, elucidating the differences between bulk and
epitaxially strained thin films in the [001] and 
[110] orientations.

In section \ref{methods} we give a short
description of our methods used in this work. Section III presents our
most important results, which we split into two parts. In part
(A) we concentrate on the structural parameters of a strain-free
structure and highlight a confusion in the literature
about the lattice parameters of \SRO. Part (B) deals with [001]
and [110] oriented structures of \SRO grown with tensile and
compressive misfit strain. Section IV presents discussions of
three selected topics which follow from our results and deserve more
attention. Conclusions are given in Section V.

\begin{figure}
\includegraphics[width=7.5cm]{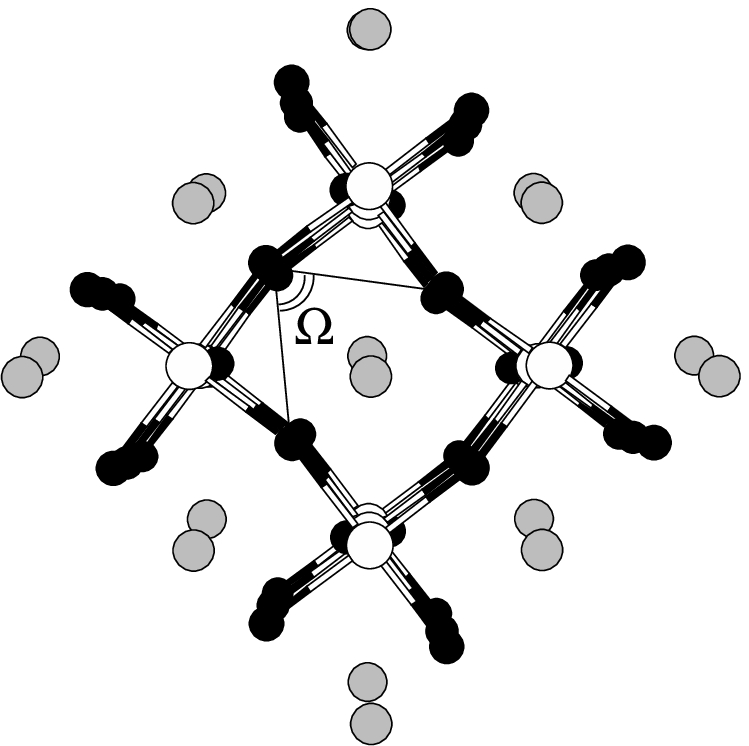}\\
\vspace{5mm}
\includegraphics[width=7.5cm]{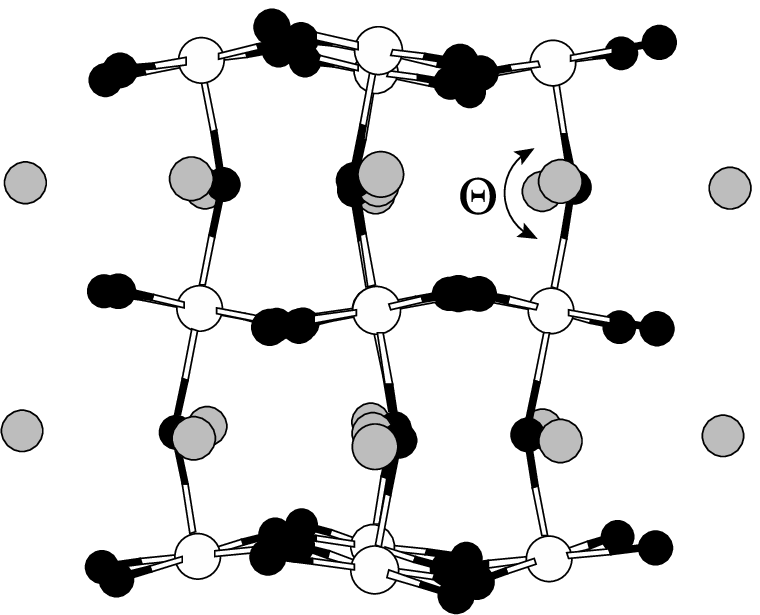}
  \caption{\label{side_top}  
(Top picture) The top view of the
  structure of \SRO showing the distortion which we refer to as a
  rotation. The corner-connected oxygen octahedra rotate in opposite
  directions around the [001] direction. This distortion changes the
  $a$ and $b$ parameters of the structure equally.
  (Bottom picture) The tilting of the \SRO structure shown projected
  along the [110] direction (or $b$ axis of the $Pbnm$ 
  structure). In this view one can see the folding of the [001]
  plane that is responsible for the shrinking of the $Pbnm$ structure
  along the $a$ axis.}
\end{figure}

\section{\label{methods}methods}

All calculations are performed with the Vienna {\it ab-initio}
Simulations Package({\sc VASP})\cite{kresse:1993,kresse:1995}, with
the Ceperley-Alder parametrization of the local spin-density
approximation (LSDA) and projector-augmented wave
potentials \cite{kresse:1999,blochl:1994}. The calculations are
performed using a plane-wave energy cutoff of 500~eV and the
convergence was checked with a cutoff of 600~eV. For 
calculations with the $Pbnm$ unit cell of \SRO, which contains 4
formula units (f.u.) or 20 atoms per cell, a $12\times12\times 
10$ Monkhorst-Pack $k$-point mesh  
is found to yield convergence of all properties computed here and is used in all
self-consistent calculations. For a simple cubic perovskite cell with
5 atoms we used a $12\times12\times12$ $k$-point sampling.

In order to isolate the effects of epitaxial strain, we
perform bulk calculations of the $Pbnm$ ground-state structure with
lattice vectors constrained to match a hypothetical substrate.
Specifically, for a series of
values of misfit strain, we fix the in-plane ($xy$ plane) lattice
vectors, allowing the $c$ lattice parameter (perpendicular to the 
substrate) and the internal ionic coordinates to fully relax within
the symmetry of the $Pbnm$ space group. The ions were free to move
until the Hellmann-Feynman forces were less than 4 meV/\AA.

Two different angles characterize the degree of
rotation and tilting of the oxygen octahedra in \SRO. 
The rotation and tilting represent basically the same type
of distortion, but with respect to different directions. It is
conventional to define the rotation along the z axis separately from those
along the x and y axes, referred to as tiltings. For what
follows, we define the dihedral angle $\Theta$ to be along the 
Ru$_1$-O$_2$-Ru$_3$ bond shown in Fig.\ref{structure}, and the 
tilting angle is given by (180-$\Theta$)/2. Similarly, the 
rotation angle is defined through the relation (90-$\Omega$)/2, where  
$\Omega$ is an angle shown in
Figures \ref{structure} and \ref{side_top}. The rotation and tilting
commute with each other and uniquely describe the network of oxygen
octahedra, assuming the octahedrons are regular. We will return to the
validity of this assumption later in 
section III \footnote{A group-theoretical analysis shows that the
  octahedra are not necessary regular. Howard and Stokes in
  Ref.\onlinecite{Howard} write ``departures from the regularity
  are allowed by the space-group symmetry and certainly
  expected," though the differences from regular octahedra should be 
  at most second order in the tilt angle. We show in this work that
  the impact of the octahedron distortions may not be neglected in \SRO.}.

\section{Results}

\subsection{Strain-free structure of $\mathbf{SrRuO_3}$}

Starting from the experimental space group $Pbnm$, we first obtain the ground
state structure of \SRO by relaxing all seven free internal structural
parameters (internal atomic coordinates) and three lattice
parameters. Our results, along with a comparison with the 
experimental values for different perovskite systems, are shown in
Tables~\ref{experiment} and \ref{theory}. 
The calculated parameters are in good agreement with experiment. 
The slight underestimate of the lattice constant is typical for LSDA. 
We calculate the FM state in $Pbnm$ structure to have an
energy 186 meV/f.u. lower than that of the cubic perovskite FM system,
which was optimized within the $Pm\overline{3}m$ symmetry (see Table
\ref{calculated}). A previous first-principles study
(Ref.~\onlinecite{singh:1996}) computed a comparable energy
difference, 140 meV/f.u., using structures with experimental lattice
parameters.

Experimental values of the magnetic moment in \SRO range between
1.1 and 1.6 $\mu_B$/f.u.; the spread in values is expected given the
difficulty of making single-domain samples and their large
magnetocrystalline anisotropy \cite{kanba:1976,klein}. According
to Ref.~\onlinecite{MazinSRO}, only approximately 60\% of the total
magnetic moment is Ru derived, while the remaining 40\% is distributed
among oxygen sublattice sites.  
A similar spread of the values of the magnetic moment in \SRO has been
reported from theoretical investigations. Usually they are 
obtained using the LSDA and experimental lattice parameters for the
orthorhombic $Pbnm$ structure. Such calculations reported 
magnetic moments ranging between 0.97 and
1.96~$\mu_B$/f.u. \cite{allen:1996,singh:1996,Santi,okamoto}

With an intermediate degree of relaxation, keeping the experimental
lattice parameters fixed, but allowing the Wyckoff positions to
change, we obtain 1.2 $\mu_B$/f.u. After a complete relaxation 
of the structure, our magnetic moment calculated with LSDA 
is 0.98 $\mu_B$/f.u., smaller than that obtained with
experimental lattice parameters, as would be expected with the smaller
LSDA lattice parameters. A summary of all calculated lattice
parameters appears in Table~\ref{theory}. The LSDA slightly
overestimates bond strength and therefore favors stronger
hybridization of the Ru-$d$ and O-$p$ states, giving smaller lattice
parameters and consequently smaller magnetic moments.

\begin{table*}
 \caption{\label{experiment}  
Experimental structural parameters for several perovskite systems with
$Pbnm$ symmetry taken from
Ref.~\onlinecite{jones:1989,Ross2004,Cockayne2000,Kiyama,LB7e,liferovich}.
We focus in this list on the relations of the $a$, $b$ and $c$ lattice
parameters. In most of the systems $a<b$ is satisfied due to the
tilting of the oxygen octahedra, whereby only the $a$ vector
shrinks, while the $b$ vector remains unchanged. From this table, one
can see that for SrRuO$_3$ the ratio $a<b$ is not satisfied. } 
\begin{tabular*}
{0.9\textwidth}
{@{\extracolsep{\fill}}l l c c l l l l}
\hline
\hline
 & $a$(\AA) & $b$(\AA) & $c$(\AA)& Wyckoff position & $x$ & $y$ & $z$
\\
\hline
SrRuO$_3$  & 5.5670 & 5.5304 & 7.8446 & Sr (4c) & -0.0027 & 0.0157 &
  0.25 \\
   & & & & Ru (4b) & \;0.5 & 0.0 & 0.0 \\ 
   & & & & O (8d) & \;0.7248 & 0.2764 & 0.0278 \\ 
   & & & & O (4c) & \;0.0532 & 0.4966 & 0.25 \\ 
GdFeO$_3$  & 5.3510 & 5.6125 & 7.6711 & Gd (4c) & -0.0015 & 0.0626 &
  0.25 \\
   & & & & Fe (4b) & \;0.5 & 0.0 & 0.0 \\ 
   & & & & O (8d) & \;0.6966 & 0.3011 & 0.0518 \\ 
   & & & & O (4c) & \;0.1009 & 0.4669 & 0.25 \\ 
GdAlO$_3$  & 5.2537 & 5.3030 & 7.4434 & Gd (4c) & -0.0079 & 0.0376 &
  0.25 \\
   & & & & Al (4b) & \;0.5 & 0.0 & 0.0 \\ 
   & & & & O (8d) & \;0.7147 & 0.2855 & 0.0387 \\ 
   & & & & O (4c) & \;0.0724 & 0.4863 & 0.25 \\ 
CaTiO$_3$  & 5.3804 & 5.4422 & 7.6417 & Ca (4c) & -0.0065 & 0.0349 &
  0.25 \\
   & & & & Ti (4b) & \;0.5 & 0.0 & 0.0 \\ 
   & & & & O (8d) & \;0.7111 & 0.2884 & 0.0372 \\ 
   & & & & O (4c) & \;0.0707 & 0.4842 & 0.25 \\ 
CaRuO$_3$  & 5.3408 & 5.5311 & 7.6460 & Ca (4c) & -0.0150 & 0.0560 &
  0.25 \\
   & & & & Ru (4b) & \;0.5 & 0.0 & 0.0 \\ 
   & & & & O (8d) & \;0.6930 & 0.2970 & 0.0530 \\ 
   & & & & O (4c) & \;0.0910 & 0.4670 & 0.25 \\ 
GdScO$_3$  & 5.4862 & 5.7499 & 7.9345 & Gd (4c) & \; -0.0163& 0.0594  &
  0.25 \\
   & & & & Sc (4b) & \;0.5 & 0.0 & 0.0 \\ 
   & & & & O (8d) & \; 0.6931 & 0.3007  & 0.0556 \\ 
   & & & & O (4c) & \; 0.1183 & 0.4465 & 0.25 \\ 
DyScO$_3$  & 5.4400 & 5.7130 & 7.887 & Dy (4c) & \; -0.0172 & 0.0607 &
  0.25 \\
   & & & & Sc (4b) & \; 0.5 & 0.0 & 0.0 \\ 
   & & & & O (8d) & \; 0.6926 & 0.3040 & 0.0608 \\ 
   & & & & O (4c) & \; 0.1196 & 0.445 & 0.25 \\ 
\hline
\hline
    \end{tabular*}
\end{table*}

\begin{table*}
    \caption{\label{theory}  
Calculated structural parameters for several perovskite systems with
$Pbnm$ symmetry. 
Theoretical results also show the distortion of SrRuO$_3$ for which
the orthorhombic distortion is opposite to other perovskites. We
introduce parameters, $\Delta a$ and $\Delta b$, which
characterize shrinking of the $a$ and $b$ lattice parameters,
respectively. The parameter $a$ shrinks due to the tilting, while $b$
shrinks due to the rectangular distortion of the horizontal middle
plane of the oxygen octahedra.} 
\begin{tabular*}
{0.9\textwidth}
{@{\extracolsep{\fill}}l l c c l l l l l l l}
\hline
\hline
 & $a$(\AA) & $b$(\AA) & $c$(\AA)& Wyckoff pos. & $x$ & $y$ & $z$
& Tilting & $\Delta a$  & $\Delta b$
\\
\hline
SrRuO$_3$ & 5.5031 & 5.4828 & 7.7546 & Sr (4c) & -0.0050 & 0.0296 &
  0.25 &  10.21$^o$  &  1.58  &  1.75 \\
   & & & & Ru (4b) & \;0.5 & 0.0 & 0.0 & & &\\ 
   & & & & O (8d) & \;0.7165 & 0.2834 & 0.0336 & & &\\ 
   & & & & O (4c) & \;0.0647 & 0.4941 & 0.25 &  & & \\
CaRuO$_3$ & 5.2090 & 5.5297 & 7.5512 & Ca (4c) & -0.0211 & 0.0645 &
  0.25 &  16.79$^o$  & 4.26  & 2.15 \\
   & & & & Ru (4b) & \;0.5 & 0.0 & 0.0  & & &\\ 
   & & & & O (8d) & \;0.6929 & 0.2996 & 0.0519 & &  &\\ 
   & & & & O (4c) & \;0.1034 & 0.4665 & 0.25  & & & \\
CaTiO$_3$ & 5.2900 & 5.4007 & 7.5334 & Ca (4c) & -0.0099 & 0.0468 &
  0.25 & 13.26$^o$  & 2.67  & 0.75 \\
   & & & & Ti (4b) & \;0.5 & 0.0 & 0.0  & & &\\ 
   & & & & O (8d) & \;0.7065 & 0.2926 & 0.0425 & & & \\ 
   & & & & O (4c) & \;0.0811 & 0.4790 & 0.25 &  &  & \\
\hline
\hline
    \end{tabular*}
\end{table*}

In addition to the ferromagnetic (FM) configuration, we 
optimize the bulk structural parameters of SRO with different
initial magnetic orderings, including the non-magnetic (NM) phase and
three different antiferromagnetic (AFM) spin arrangements. 
Several local minima in the total energy are
found. Among these minima, the FM configuration is computed to be the
most stable state, in agreement with experiment. The $A$-type AFM and
$C$-type AFM, and NM configurations are found to be 6.77meV/f.u.,
6.44meV/f.u., and 8.15meV/f.u. higher in energy than the FM ground
state, respectively. 
The $G$-type AFM configuration is found to be unstable: when
relaxed, it reverts to the NM minimum.  
In the remaining of this paper, we focus on the FM state and its
comparison with the NM state.

At this point, we turn to a discussion of the lattice parameters of 
the $Pbnm$ structure. 
From Tables \ref{experiment} and \ref{theory}, we can see that most of
the $Pbnm$ GdFeO$_3$-type structures have $a < b < c$. Interestingly,
however, in the case of \SRO this relation is not satisfied.
\SRO is the only compound we know of for which this is the case. 
This has led to a confusion
in the literature where one can find reports both of $a > b$ (see
Refs.\onlinecite{jones:1989,Feizhou,zakharov,gausepohl,vasco}) and  
$a < b$ (see Refs.\onlinecite{gan1999,kennedy,klein,Santi}); in the
latter case parameters are apparently switched to be
consistent with overall trend in the  GdFeO$_3$-type
structures. To clarify the unique situation in \SRO other 
geometrical factors making up the orthorhombic
distortions of the $Pbnm$ structure need to be considered, as we now
explain.

\begin{figure}
\includegraphics*[angle=0,width=9.0cm]{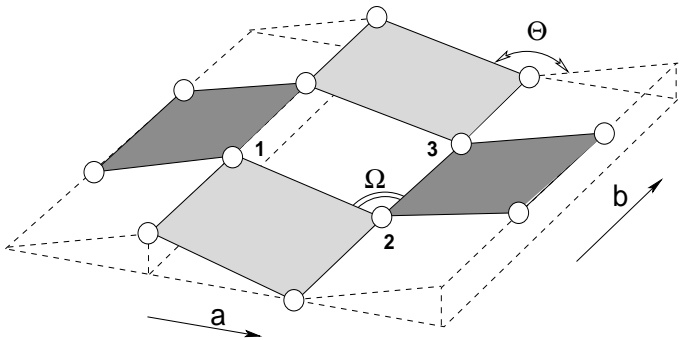}
  \caption{\label{folding}
This picture illustrates schematically the fact that as a
result of tilting the $Pbnm$ structure is necessarily
orthorhombic. This appears from simple geometrical considerations when
tiltings $a^-a^-$ have the effect of folding of the Ru-O (001) atomic
planes. If the folding develops along the $a$ axis, while another
lattice vector $b$ remains unchanged, we have $a < b$.      
The importance of this point can be seen when the $Pbnm$ structure is
subject to the epitaxial constraints enforcing $a =
b$. The rotation around the $z$ axis is different because it
changes the $a$ and $b$ parameters of the $Pbnm$ structure in equal
proportions. So, the $a=b$ constraint requires the middle plane of the
oxygen octahedra to shrink along the $b$ axis.
In order to characterize this epitaxially constrained structure we
introduce the notation e-$Pbnm$. The angles shown define degrees of tilting
and rotation as discussed in the section ``Methods''.} 
\end{figure}

Figure \ref{side_top}, together with a sketch in  Fig.\ref{folding},
illustrates the relationship between octahedral tilting and rotation
and the orientations of the interatomic bonds.  
The in-plane lattice parameters of the $Pbnm$ structure are not equal,
that is $a \ne b$, as a consequence of the tilting; the rotation on the
other hand does not change the ratio of $a$ to $b$.  
For arrangements with $Pbnm$ symmetry, if the octahedra were
regular, tilting would require $a < b$, and not the other way
around. This simple 
geometrical picture explains the orthorhombic shape of most of the
known perovskite systems listed in Table \ref{experiment}, with the
notable exception of \SRO. 

Our theoretical calculations confirm that for
\SRO,  $a > b$ (see  Tab.~\ref{theory}), while for the systems
like CaRuO$_3$ and CaTiO$_3$,  $a < b$. 
An analysis of the results of our calculations shows that the reason
for the apparent anomaly in \SRO is that the oxygen octahedra 
are not regular: the ``horizontal'' middle plane is not
square, but rather is rectangular. The rectangular distortion occurs
along the $b$ axis partially compensating the orthorhombic
distortion originating with the tilting. Thus, two different
contributions determine the orthorhombic distortion (the $a/b$ ratio)
in the $Pbnm$ structure. 

In Table~\ref{theory} we summarize the impact of distortions
coming from each contribution. The
anomalous orthorhombic distortion in \SRO is 
derived from geometrical considerations and those parameters which we
already have: 
$a$, $b$, $c$, tilting and rotation angles. If
$a^{\mathrm{oct}}$ and $b^{\mathrm{oct}}$ describe the equatorial
plane of the octahedra, then the lattice parameters of $Pbnm$ 
can be expressed as follows:

\begin{equation}
\label{eqone}
\begin{cases}
& a = 2 a^{\mathrm{oct}} \cos{\mathrm{(rotation)}}
  \cos{\mathrm{(tilting)}}; \\
& b = 2 b^{\mathrm{oct}} \cos{\mathrm{(rotation)}}. 
\end{cases}
\end{equation}

Consequently, the orthorhombic distortion can be expressed as:

\begin{equation}
\label{atob}
 \frac{a}{b} = \left[\frac{a^{\mathrm{oct}}}{b^{\mathrm{oct}}}\right]
  \cos{\mathrm{(tilting)}}.
\end{equation}

These simple relations show that the internal distortions 
can be readily determined from fundamental parameters of
the $Pbnm$ structure. However, it is important to note that the oxygen
octahedra are not necessarily regular, i.e. $a^{\mathrm{oct}}$ and
$b^{\mathrm{oct}}$ are free parameters unique for each
material.

\subsection{Epitaxially constrained structure: \boldmath  e-$Pbnm$}

We now consider the structure and properties of
epitaxially strained $Pbnm$ \SRO. Experimentally, \SRO films can
grow with either [001] or [110] orientations on various substrates
like LaAlO$_3$, SrTiO$_3$, DyScO$_3$ or GdScO$_3$
\cite{vasco,gan1999} (see Ref.~\onlinecite{kennedy} for good
illustrations of the two geometries). The [001] case is a relatively 
straightforward extension of 
previously studied perovskite multi-layers \cite{Karen,jeff2}, except
that additional degrees of freedom must be considered to account for
the tilting and rotation. 
In the [110] orientation, the situation is quite
different, as we discuss in detail below.

\subsubsection{Geometries of the [001] and [110] oriented films}

For the [001] oriented films, the $a$ and $b$ parameters of the 
orthorhombic $Pbnm$ lattice are constrained to be equal (in order to
match the square lattice of the substrate). We refer to these
constrained structures, with all other structural parameters relaxed,
as e-$Pbnm$[001], where ``e'' indicates ``epitaxial''. 
Misfit strain is measured relative to the lowest energy e-$Pbnm$[001]
structure, the computed parameters being given in
Table \ref{calculated}. Note that the zero misfit strain structure is
not the same as the strain-free structure discussed in the previous
subsection, as the latter is not compatible with the epitaxial
constraints. For the e-$Pbnm$ structure we have $\Delta a =
\Delta b$ (from Table \ref{theory}), i.e. two kinds of distortions
compensate each other, as they are forced to by the imposed conditions.

\begin{figure}
\includegraphics*[angle=0,width=7.5cm]{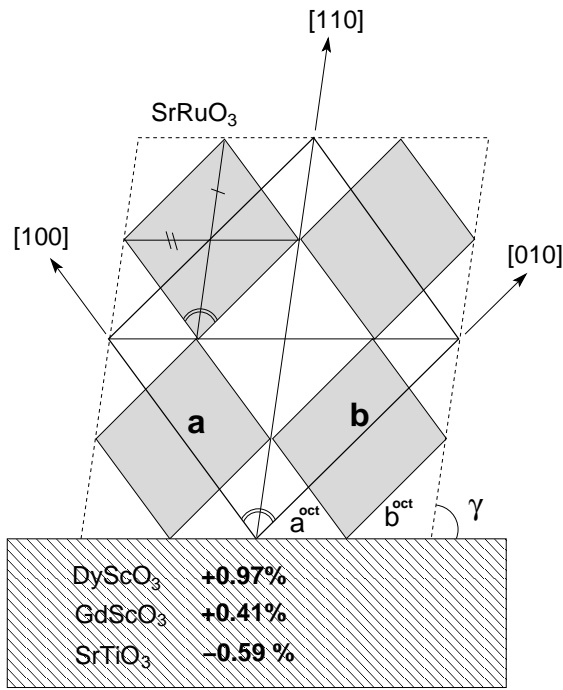}
  \caption{\label{monoc}
Schematic presentation of the epitaxial structure of \SRO grown in the
[110] orientation. The grey parallelograms represent the equatorial
planes of the oxygen octahedra in the (001) plane.
This geometry implies that at zero in-plane strain
the structure still have the orthorhombic symmetry, but its [110]
direction is not normal to the substrate. Due to the
orthorhombic symmetry, there is an angle $\gamma$ defined by
eq.\eqref{diag}. Applying in-plane strain changes this angle, but at
the same time the orthorhombic symmetry is broken. \SRO grown with the  
[110] orientation under strain is monoclinic.} 
\end{figure}

In the [110] orientation, we consider a general case when the
\SRO film is grown on an orthorhombic substrate. We assume that our
substrate has in-plane non-equal lattice parameters, but they are kept
orthogonal. If we made the in-plane parameters equal, this would
correspond, for example, to a cubic lattice of SrTiO$_3$ with a
compressive lattice mismatch of about 0.59\%. But making the in-plane
lattice rectangular brings us close to the situation with
GdScO$_3$ or DyScO$_3$ substrates, whose structures match
nicely with [110] oriented structure of \SRO (we refer to it as
e-$Pbnm$[110], although we show that it may have lower symmetry). In 
this case \SRO is subject to substantially different constraints 
compared with the [001] orientation. The diagonal $\sqrt{a^2+b^2}$ is
fixed and the lattice parameter $c$ is the in-plane
parameter. The out-of-plane lattice vector   
is not perpendicular to the surface (see Figure \ref{monoc}), because
its angle with the normal to the surface is given by the
difference between parameters of $Pbnm$, $a$ and $b$:
   
\begin{equation}
\label{diag}
 \gamma = 2 \arctan{\left( \frac{b}{a} \right)} = 2 \arctan{\left(
 \frac{b^{\mathrm{oct}}}{a^{\mathrm{oct}}}\cos{\mathrm{(tilting)}} \right)}
\end{equation}

Although the diagonal $\sqrt{a^2+b^2}$ is fixed on the substrate, the
ratio $a/b$ is free to change. From eq.\ref{eqone} we see that
$b/a=(b^{\mathrm{oct}}/a^{\mathrm{oct}}) \cos{\mathrm{(tilting)}}$. 
At zero in-plane strain the [110]-oriented \SRO retains the bulk
structure if it is grown on SrTiO$_3$, but not in the case of
[001] oriented GdScO$_3$ or DyScO$_3$ substrates.

\subsubsection{ Structural changes as a function of strain}

\begin{table*}
    \caption{\label{calculated}  
      Computed structural parameters of \SRO in the cubic
      perovskite (space group $Pm\bar{3}m$) and  the epitaxially
      constrained structure e-$Pbnm$[001] in the zero 
     strain state. Results for both FM and NM configurations are listed in
     order to show that they do not have significant
     differences (later on, it will be shown that applying epitaxial
     strain yields important differences between the FM and NM cases).
     The Wyckoff positions for all structures refer to space group
     $Pbnm$.}
\begin{tabular*}
{0.9\textwidth}
{@{\extracolsep{\fill}}l l c c l l l l l l l}
\hline
\hline
 & $a$(\AA) & $b$(\AA) & $c$(\AA)& Wyckoff pos. & $x$ & $y$ & $z$ 
& Tilting & $\Delta a$  & $\Delta b$
\\
\hline
  e-$Pbnm$(FM) & 5.4929 & 5.4929 & 7.7514 & Sr (4c) & -0.0055 & 0.0303
  & 0.25 & 10.48$^o$ & 1.67\%  & 1.67\% \\ 
   & & & & Ru (4b) & \;0.5 & 0.0 & 0.0 & & & \\ 
   & & & & O (8d) & \;0.7160 & 0.2837 & 0.0336 & & & \\ 
   & & & & O (4c) & \;0.0650 & 0.4937 & 0.25 & & & \\
  e-$Pbnm$(NM) & 5.4929 & 5.4929 & 7.7440 & Sr (4c) & -0.0051 & 0.0304
  & 0.25 & 10.48$^o$  &  1.67\% &  1.69\% \\ 
   & & & & Ru (4b) & \;0.5 & 0.0 & 0.0 & & &\\ 
   & & & & O (8d) & \;0.7155 & 0.2842 & 0.0345 & & &\\ 
   & & & & O (4c) & \;0.0650 & 0.4945 & 0.25 & & & \\
  Cubic(FM) & 5.5070 & 5.5070 & 7.7880 & Sr (4c) & \;0.0 & 0.0 & 0.25
& 0.0 & 0.0 & 0.0 \\ 
   & & & & Ru (4b) & \; 0.5 & 0.0 & 0.0 & & &  \\ 
   & & & & O (8d) & \; 0.75 & 0.25 & 0.0  & & & \\ 
   & & & & O (4c) & \; 0.0 & 0.5 & 0.25  & & & \\
\hline
\hline
    \end{tabular*}
\end{table*}

\begin{figure}
\includegraphics*[width=7.5cm]{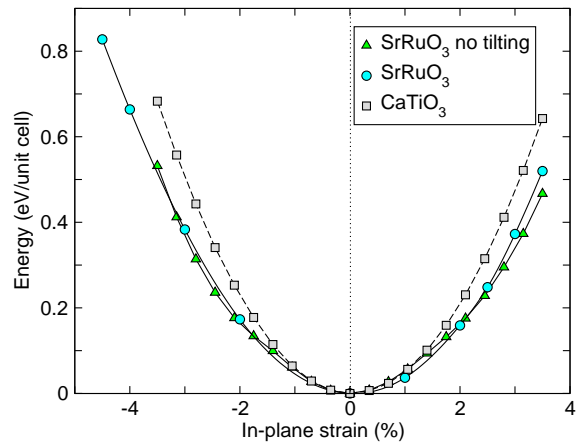}
  \caption{\label{totenstr}  
The comparison of the elastic energies of FM e-$Pbnm$[001] \SRO and NM
e-$Pbnm$[001] CaTiO$_3$ as a function of the applied misfit
strain. The in-plane lattice parameters were fixed by the value of the 
misfit strain, while all other parameters of the structures were relaxed to
the zero-force states. In addition we show here the total energy of
the simple $P4/mmm$ perovskite \SRO when rotation and tilting are not
allowed. This allows us to conclude that the tilting does not change
the elasticity of the structure with respect to the misfit strain.}
\end{figure}

\begin{figure}
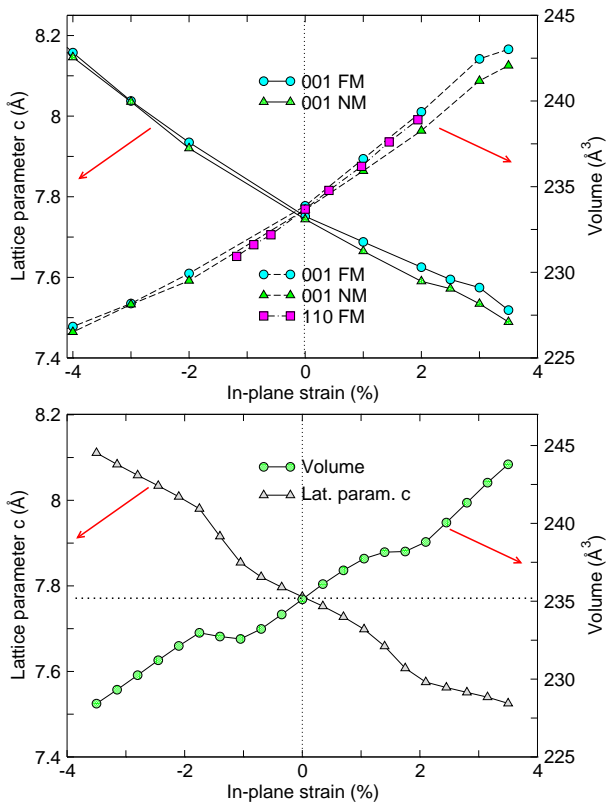

\includegraphics*[width=8.0cm]{figure9.eps}\\
\includegraphics*[width=8cm]{figure10.eps}
  \caption{\label{latticech}
Lattice parameter $c$ and the volume of the [001] and [110] oriented
perovskite structures of \SRO under epitaxial strain.
(Top)
 The epitaxially constrained e-$Pbnm$ structure. Both
 volume change and tetragonal distortion cost energy, whereby the
 structure changes these degrees of freedom simultaneously. 
 Both [001] and [110] oriented films show similar changes in the
 volume as a function of the strain.\\  
(Bottom)
Non-tilted [001] oriented simple perovskite $P4/mmm$.
Both curves show remarkable difference of their
behavior as compared to the e-$Pbnm$ structure. A comparison to the 
Fig. \ref{magch} shows that the distortions of the $P4/mmm$ are
strongly influenced by the magnetoelastic coupling.
}
\end{figure}

As the epitaxial constraint is varied, the shape of
the cell is changed, which increases its internal energy. In
Fig.~\ref{totenstr}, we plot the elastic energy of ferromagnetic \SRO for
misfit strains ranging from -4.5\% to +3.5\%. Surprisingly, the
energetic penalties for the in-plane distortion are nearly the same
for both e-$Pbnm$ and $P4/mmm$ structures, though the latter has
no internal degrees of freedom to relax and reduce the energy.
Moreover, comparing these curves with another
perovskite, CaTiO$_3$, we see that their behavior is
strikingly similar. Changes of the volume are shown in
Fig.~\ref{latticech}.

\begin{figure}
\includegraphics*[width=8cm]{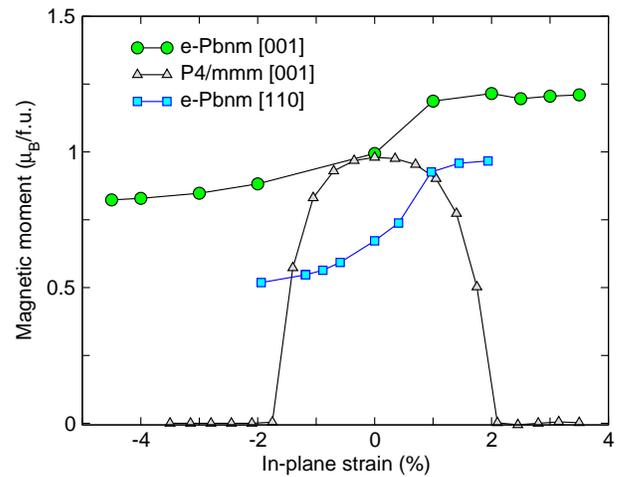}
  \caption{\label{magch} 
   The total magnetic moment of \SRO as a function of
   strain. In the case when tilting is not
   allowed ($P4/mmm$), the deformation of the Ru-O octahedra, occurring
   due to the strain, becomes so strong that the total magnetic moment
   drops down to zero.  
   Redistribution of the electron orbitals changes the magnetic
   interactions. Beyond critical values of the strain,
   \SRO becomes non-magnetic.  
   When the tilting is present ($Pbnm$ structure) 
   the magnetic moment is retained even at large in-plane strain. In the
   [110] oriented film the magnetic moment is significantly smaller,
   which we attribute to the geometry of the oxygen cages and the
   corresponding changes in the electronic structure (see DOS in
   Fig. \ref{dos_one}).}
\end{figure}

\begin{figure}
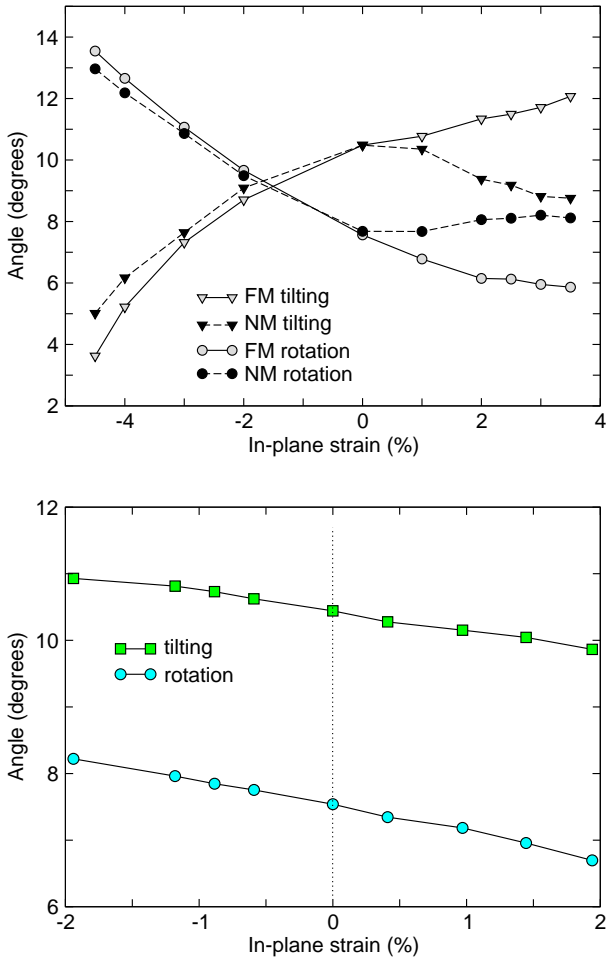

\includegraphics*[width=8cm]{figure12.eps}\\
\vspace{5mm}
\includegraphics*[width=8cm]{figure13.eps}
  \caption{\label{ang}  
  Tilting and rotation angles of the oxygen octahedra as a function
  of strain.
(Top)
The e-$Pbnm$[001] structure shows a sudden change in the behavior of
  the tilting and rotation angles in the NM system, which turns out to be
  due to the additional contraction of the oxygen octahedra along the z
  axis (see  Fig.\ref{bond}). Both in the FM and NM cases under tensile
  strain the rotation angles stay close to a value of 10
  degrees.   
(Bottom)
In the e-$Pbnm$[110] orientation, both angles are fixed by the
  substrate constraint.
  }
\end{figure}

The degree of tilting and rotation of the oxygen octahedra reflects
the impact of epitaxial strain. 
Figure \ref{ang} shows the evolution of these angles for both [001]
and [110] orientations. For [110] the two angles decrease together.
On the other hand, for the [001] orientation, the angles behave oppositely:
for increasing compressive or
tensile strain, the rotation or tilting become more pronounced,
respectively. However, we find that the angles exhibit surprising
behavior under tensile strain in the NM~state (discussed further
below).

\begin{figure}
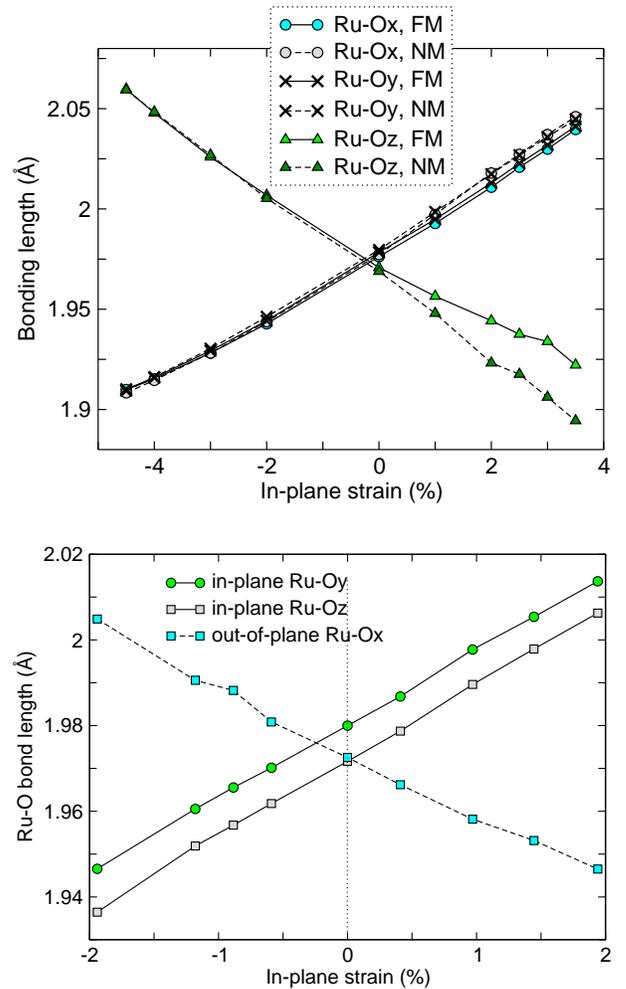

\includegraphics*[width=8cm]{figure14.eps}\\
\vspace{5mm}
\includegraphics*[width=8cm]{figure15.eps}
  \caption{\label{bond} 
   Bond lengths of Ru-O inside the oxygen octahedral cages as
   functions of in-plane strain. 
(Top)
 The [001] oriented film with FM and NM configurations. 
 Under the tensile strain the NM and FM curves split, which is due to
 the additional contraction of the oxygen octahedra along the z axis in
 the NM case, which is reflected in smaller tilting angles shown in
 Fig.\ref{ang}.  
(Bottom)
 In the [110] orientation, the substrate has an orthorhombic shape,
   therefore the in-plane parameters of the \SRO film are
   distorted. In order to be consistent we use the notations of the
   [001] orientation. Therefore, the x axis in this case is
   out of plane. 
}
\end{figure}

In addition to the orientation of the oxygen
octahedra, we observe their shape is also changing. This is especially
important because we have to understand the behavior of the angles in
the NM case shown in Figure \ref{ang}. 
The splitting of two Ru-Oz curves in Figure \ref{bond} can be
explained by the fact that the
NM configuration allows for additional tetragonal contraction of the
oxygen octahedra along the $z$ axis. This contraction manifests itself in 
the trend in angles shown in Fig.\ref{ang}. 
Connection of this effect with the magnetic properties of \SRO will be
discussed later in Section IV.

Applying strain induces a change in volume. 
In the [110] orientation, the volume can be modified by  
rotating the angle $\gamma$ or by expanding (or contracting) the
out-of-plane lattice parameter. The latter can be achieved only if the
symmetry of the oxygen octahedra is  broken. Namely, the
middle plane Ru-O bonds would be required to have different length; in
the case of [001] oriented films they are the same (see
Fig. \ref{bond}). This additional lowering of the symmetry alters the 
electronic environment of the Ru ion leading to a drop of its magnetic
moment, as shown in Fig.\ref{magch}. The [110] oriented structure of
\SRO becomes monoclinic \footnote{We continue to refer to this
structure as e-$Pbnm$[110] in order to avoid introducing new
notation, however this structure has lower symmetry if the in-plane
strain is applied.} (see Fig.~\ref{monoc}).

\section{\label{discussions}discussions}

From the results reported above, we selected three particular points for further
discussion. Our focus here is on the magnetostructural coupling in \SRO
and its dependence on the film orientation.    

 \subsection{Magnetoelastic coupling}

Fig.~\ref{latticech} exhibits some nonlinearity in the
volume of the $P4/mmm$ with misfit strain, in contrast with
the e-$Pbnm$[001] structure. To understand the behavior of $P4/mmm$ 
it is useful to compare Figures \ref{latticech} and \ref{magch}.
Both curves show inflection points that in Fig.\ref{latticech} coincide
with abrupt drops in magnetic moment. 
Evidently, the magnetic moment, via the spin configuration of the Ru
ion, stabilizes a symmetrical non-deformed shape of the oxygen
octahedra. 
Concurrently, the in-plane strain leads to the out-of-plane
 lattice expansion (contraction) under the compressive (tensile)
 strain. At some critical strain the contraction becomes more
 favorable than the magnetic orbital ordering and the lattice 
 distorts. On the other hand, in the e-$Pbnm$ structure, the tilting
 and rotation preserve the shape of octahedra and
 the magnetic orbital ordering survives.
   
The behavior described here suggests
a strong magnetostructural interaction, which may potentially explain
past experiments showing the anomalous thermal expansion of \SRO
\cite{Kiyama}. We reserve a more complete investigation for future work.

\subsection{Anomaly of tilting and FM-NM transition}

In the previous section we saw the effect of coupling between the
distortions of the oxygen octahedra and the stability of the magnetic 
moment on Ru in the $P4/mmm$ symmetry. This allows us to understand
what we see in the case of e-$Pbnm$[001] structure.   
In Figure \ref{bond}, we see an extra contraction of the Ru-Oz bond,
which occurs in the NM phase under tensile strain.
In contrast, in the FM phase the oxygen octahedra keep their
shape by monotonically increasing the tilting angle.  

This difference in the behavior of the magnetic and non-magnetic
phases can be explained from a simple spin configuration argument known for
\SRO \cite{Itoh}. If the structure is not distorted, four $4d$
electrons of Ru$^{4+}$ occupy 3-fold degenerate $t_{2g}$ orbitals
leading to ($\uparrow\!\!t^3_{2g}, \downarrow\!\!t^1_{2g}$), whereby
two spins are not compensated and the Ru ion has a large magnetic
moment. If the oxygen octahedra are contracted along some of the Ru-O
bonds, the $t_{2g}$ orbitals split producing a non-magnetic spin
configuration for Ru,  $\uparrow\!\!t^2_{2g},
\downarrow\!\!t^2_{2g}$. 

Such a transition is possible and can have significant implications
for the magnetostructural tuning of the \SRO based
heterostructures. 
One could change the magnetic ground state  
of \SRO by using uniaxial compression. 
Starting from the FM state, one can apply a
uniaxial stress along the $c$ axis. The NM state which has a shorter
lattice parameter $c$ can become the ground state.
Some doping of the \SRO structure with larger atoms, like Ba, should
make tilting weaker and therefore favor the NM state, i.e. reduce the
amount of pressure required to transform the structure. 
We will discuss this issue in more detail with possible applications
in our next paper.

\subsection{Orientation dependence}

We now comment on the differences between [001]- and [110]-oriented
films. For the [001]-oriented films, the oxygen octahedra are free to
rotate and tilt adopting to the changing lattice parameters. In
contrast, in the [110] orientation both angles are fixed to the
parameters of substrate. Moreover, we stressed above that the [110]
oriented \SRO becomes monoclinic.
This difference has consequences for the calculated electronic structure. 
We show the electronic DOS projected on the Ru sites for four different
cases. These four plots are intended to illustrate how the electronic
environment of Ru changes when distortions of the oxygen cage network
occur.

\begin{figure}
\includegraphics*[width=8.0cm]{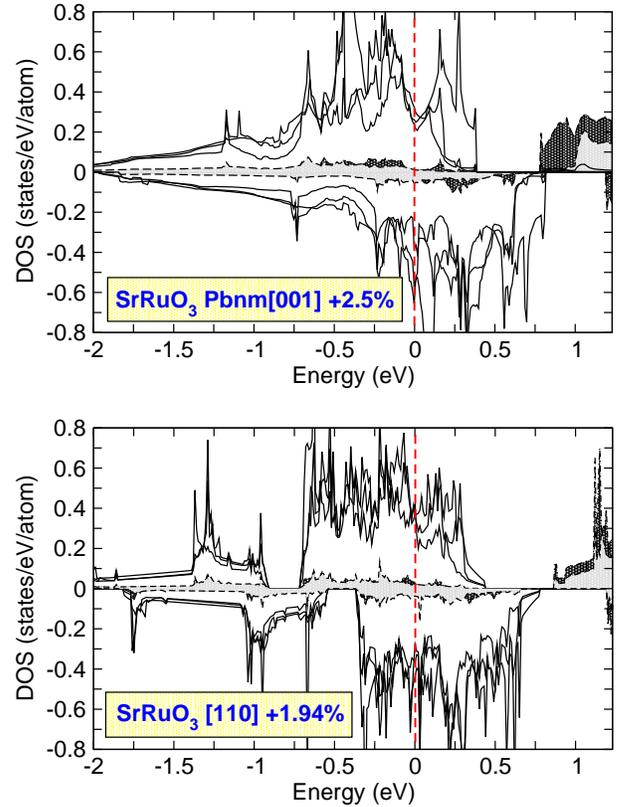}
  \caption{\label{dos_one}  
Ru site projected partial DOS of \SRO in the vicinity of $E_F$. The
electronic states with 
$e_g$ symmetry are filled with gray color. Different orientations of the
t$_{2g}$ and e$_g$ orbitals are plotted together without
distinction in order to focus on the gaps which open when distortions
are applied to the structure.
(Top) DOS for the e-$Pbnm$[001] structure of \SRO under  
  tensile strain of 2.5\%. The tensile strain reduces in-plane
  distortion of the oxygen octahedra, whereby the gap becomes
  smaller. 
(Bottom)
 DOS for the e-$Pbnm$[110] structure. The strain is now also tensile,
 but 1.94\%. There is another gap which opens about 0.8 eV (0.4 eV for
 spin-down) below the
 Fermi level. It happens because the [110] orientation required
 additional distortion of the oxygen cages, forcing the $t_{2g}$
 electrons of Ru to split.}
\end{figure}

The additional lowering of the symmetry in the [110]
orientation leads to a splitting of the t$_{2g}$ bands associated with
Ru $d$ level.
and a smaller magnetic moment (Fig.\ref{magch}). The gap opens
at about 0.8 eV below E$_F$ for the spin-up states, and 0.4 eV below
E$_F$ for the spin-down component (Fig.~\ref{dos_one}(Bottom)).

\begin{figure}
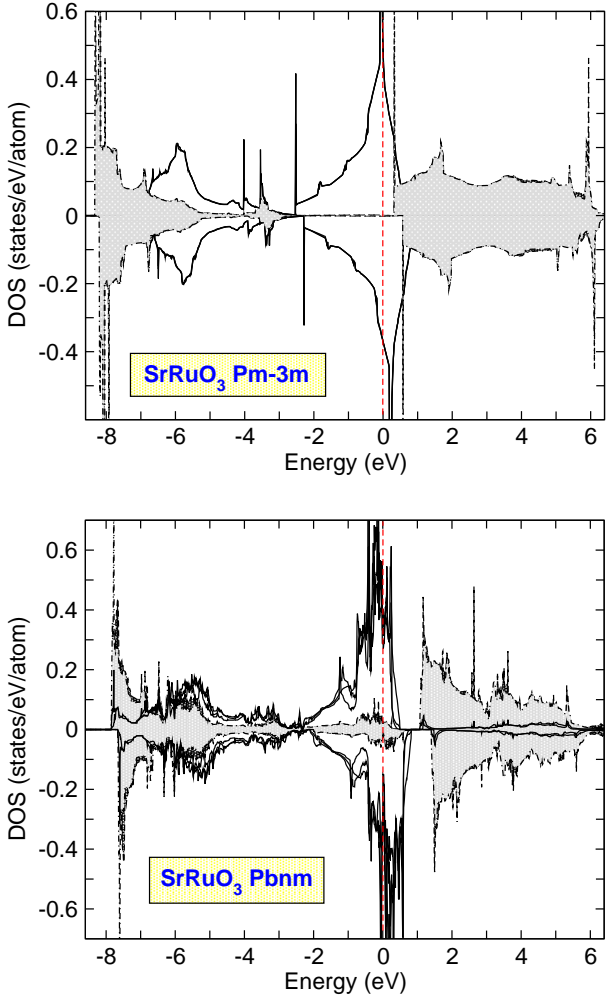

\includegraphics*[width=8.0cm]{figure16.eps}\\ \vspace{5mm}
\includegraphics*[width=8.0cm]{figure17.eps}
  \caption{\label{Rudos}  
Ru site projected partial DOS of \SRO. 
The electronic states with $e_g$
symmetry are filled with gray color. Different orientations of the
t$_{2g}$ and e$_g$ orbitals are plotted together without
distinction in order to focus on the gaps which open when distortions
are applied to the structure. (Top) DOS of the simple perovskite
structure with $Pm\overline{3}m$ symmetry. There is no gap 
above the Fermi level. 
(Bottom) 
DOS of the strain free bulk $Pbnm$ structure. There is a gap above the
  Fermi level, which opens, as we discuss in the text, due to the
  rotations of the oxygen octahedra. In Fig. \ref{dos_one} it
  will be shown that in the [110] oriented structure, one more gap
  opens.}
\end{figure}

From examination of the electronic structure in Fig. \ref{Rudos} and
\ref{dos_one} four additional features emerge. We summarize them as follows:
First, for $Pm\overline{3}m$, there are no distortions of the oxygen
octahedra, and thus there is no gap in DOS (see Fig. \ref{Rudos}(Top)); 
second, for $Pbnm$ the octahedra are rotated, thus the t$_{2g}$ and
e$_g$ bands start to ``feel'' each other and repel,
and as a consequence, there is a gap just above E$_F$ in
the DOS (see Figure \ref{Rudos}(Bottom)); 
for e-$Pbnm$[001] under tensile strain, the rotation is reduced 
and the t$_{2g}$($xy$) and e$_g$($x^2-y^2$) recover their symmetry,
whereby there repulsion becomes weaker, 
reducing the gap above E$_F$ (see Fig.~\ref{dos_one}(Top)).    
Finally, for e-$Pbnm$[110], this orientation requires further lowering
of the symmetry.  The crystal field splits the t$_{2g}$ bands
and opens an additional gap 0.8 eV (0.4 eV for spin-down) below
the $E_F$. 
%

%
%

\section{Conclusions}

In this paper, we studied properties of \SRO under conditions of
epitaxial strain and different thin film orientations and considered
how the changes that take place in \SRO may be used to
tune properties of \SRO and \SRO based heterostructures.
The results can be summarized as follows: 

We explained the mechanism leading to the orthorhombic distortion of
$Pbnm$, namely, the
tilting and the contraction of the oxygen octahedra.
The ``anomalous'' orthorhombic shape of \SRO
can be easily understood from simple geometrical considerations.

Two different orientations of the \SRO epitaxial films, e-$Pbnm$[001]
and e-$Pbnm$[110], exhibit significant differences in their structural 
properties. The latter, unless there is no in-plane strain, 
has monoclinic symmetry, while the [001] oriented film remains
orthorhombic at any reasonable value of the strain.

\SRO exhibits significant magnetostructural coupling. 
 Tensile strain reveals that in the non-magnetic state there is a
 distortion of Ru-Oz bonds which can be attributed to the change of Ru
 spin configuration, explaining the fact that this distortion does not
 appear in the magnetic state. The tilting and rotation help to
 preserve the  shape of the octahedra when epitaxial strain is
 applied, whereby the magnetic orbital ordering is preserved. By
 suppressing the tilting under strain one could obtain a non-magnetic
 state of \SRO.

\section{Acknowledgements}

This work was supported by NSF (MRSEC DMR-00-80008) and DOE grant
DE-FG02-01ER45937.
J.B.N. would like to acknowledge support by the U.S. Department of
Energy under Contract No. DE-AC02-05CH11231.
We thank E. Liskova, C. J. Fennie, D. Singh, D. Vanderbilt, M. H.
Cohen, and D. R. Hamann for their help and discussions.
We also thank P. Woodward for helpful feedback and for providing his code POTATO.


\bibliographystyle{prsty}

\end{document}